\documentclass[11pt,a4paper]{article}
\pdfoutput=1

\usepackage{jcappub}

\usepackage[english]{babel}
\usepackage{comment}
\usepackage{graphicx,rotating,amsmath,amsfonts,amssymb,multicol}
\usepackage{placeins}

\newcommand{\be}{\begin{equation}}
\newcommand{\ee}{\end{equation}}
\newcommand{\bc}{\begin{center}}
\newcommand{\ec}{\end{center}}

\newcommand{\PRexp}{2.14 \pm 0.05}

\newcommand{\gsim}{\lower.7ex\hbox{$\;\stackrel{\textstyle>}{\sim}\;$}}
\newcommand{\lsim}{\lower.7ex\hbox{$\;\stackrel{\textstyle<}{\sim}\;$}}

\newcommand{\bt}{\beta}

\newcommand{\la}{\lambda}

\newcommand{\sa}{\sigma}

\newcommand{\cL}{\mathcal{L}}

\newcommand{\MP}{\bar{M}_\text{Pl}}
\newcommand{\nmc}{\xi}
\newcommand{\pd}{\partial}

\newcommand{\TRH}{T_{\rm RH}}

\title{Minimal but non-minimal inflation and electroweak symmetry breaking}

\author[a,b]{Luca Marzola}
\author[a]{Antonio Racioppi}

\affiliation[a]{National Institute of Chemical Physics and Biophysics, R\"avala 10, 10143 Tallinn, Estonia}
\affiliation[b]{Institute of Physics, University of Tartu, Ravila 14c, 50411 Tartu, Estonia}

\emailAdd{luca.marzola@ut.ee}
\emailAdd{antonio.racioppi@kbfi.ee}

\abstract
{
We consider the most minimal scale invariant extension of the standard model that allows for successful radiative electroweak symmetry breaking and inflation. The framework involves an extra scalar singlet, that plays the r\^ole of the inflaton, and is compatibile with current experimental bounds owing to the non-minimal coupling of the latter to gravity.
This inflationary scenario predicts a very low tensor-to-scalar ratio $r \approx 10^{-3}$, typical of Higgs-inflation models, but in contrast yields a scalar spectral index $n_s \simeq 0.97$ which departs from the Starobinsky limit. We briefly discuss the collider phenomenology of the framework.
}

\arxivnumber{1606.06887}

\begin{document}
\maketitle

\section{Introduction}
\label{sec:in}
According to the present knowledge, our Universe very likely underwent a period of accelerated expansion in the first instants of its existence \cite{Guth:1980zm}. This process, known as inflation, came to an end by `reheating' our Universe, filling it up with the matter and radiation \cite{Linde:1981mu} described, at least in part, by the Standard Model (SM). The essence of inflationary dynamics can be reproduced by a scalar field, the inflaton $\phi$, which tracks its potential with negligible kinetic energy. The accelerated expansion of the Universe is then guaranteed as long as the inflaton potential is the dominant contribution into the energy balance of the latter and the slow-roll condition is satisfied.
Scalar field inflation has been studied in great detail \cite{Olive:1989nu,Lyth:1998xn,Kinney:2009vz,Tenkanen:2016idg}, including possible non-minimal interactions with gravity and within frameworks such as that of classical scale invariant (CSI) theories~\cite{Heikinheimo:2013fta,Gabrielli:2013hma}.
This last class of models 
is  particularly minimalistic given that classical scale invariance restricts the tree-level inflationary potential to the form
$V(\phi)=\frac{1}{4} \lambda_\phi \phi^4$.
Current measurements of the scalar perturbations amplitude $A_s$, however, impose $\lambda \approx 10^{-13}$ and erroneously seem to rule out the model for the resulting large value of the tensor-to-scalar ratio $r$ \cite{Ade:2015fwj}. In fact, this na\"ive picture forgets the quantum corrections of the inflationary potential caused by the SM (or BSM) particles \cite{Kannike:2014mia,Kannike:2015apa,Kannike:2015kda,Marzola:2015xbh,Kannike:2016jfs}, which are necessarily present if the reheating dynamics is to proceed via the decay of the inflaton. Given the typical smallness of the tree-level inflaton quartic coupling, quantum corrections may noticeably alter the shape of the potential especially at large field values. The most minimal CSI model, where the inflaton interacts only with the Higgs boson ($h$), is  however ruled out because the possible choices for the vacuum expectation value (VEV) of $\phi$, lead either to a too small scalar spectral index $n_s$ or to a too large Higgs mass (see \cite{Kannike:2014mia} and replace $\eta$ with $h$). Yet, it is still possible to introduce non-minimal interactions between the inflaton and gravity, in the same fashion as within Higgs-inflation models \cite{Bezrukov:2007ep}. In this case it is expected that the non-minimal coupling flatten the potential at large field values, consequently forcing the inflationary solutions to converge to the Starobinsky ones \cite{Starobinsky:1980te}. In this letter, based on the preliminary analysis in \cite{Marzola:2015xbh}, we demonstrate that this is not necessarily the case: the mentioned quantum corrections within CSI models yield solutions which depart from the Starobinsky limit and that allow for the electroweak symmetry breaking (EWSB) via the Coleman-Weinberg mechanism \cite{Coleman:1973jx}.

\section{Non-minimal CSI inflation}
\label{sec:cwi}
Before turning to the EWSB mechanism, we briefly review the non-minimal inflation scenario first sketched in \cite{Marzola:2015xbh}.
We take an inflaton sector augmented with a non-minimally coupling to gravity $\nmc$
\begin{equation}
  \cL_\phi = - \frac{\nmc \phi^2+M^2}{2}R + \frac{(\pd\phi)^2}{2} - V(\phi) \, ,
  \label{eq:fullL}
\end{equation}
being $\phi$ the inflaton and $R$ the Ricci scalar. As we are interested in configurations where the VEV of the inflaton, $v_\phi$,  is much smaller than the reduced Planck mass, $\MP=2.4\times10^{18}$~GeV, we simply set $M \simeq M_\text{Pl}$. Assuming a CSI particle content, the 1-loop effective potential
of the inflaton can be approximated in a model independent way as:
\begin{equation}
  V(\phi) = \frac{1}{4} \la_\phi(\phi)\phi^4
  \label{eq:Veff}
\end{equation}
where
\begin{equation}
  \la_\phi(\phi) = \la_\phi(\MP) \left[1 + \delta(\MP) \ln\left(\frac{\phi}{\MP}\right)\right]
  \label{eq:lrun}
\end{equation}
and $\delta=\bt_{\la_\phi}/\la_\phi$ being $\bt_{\la_\phi}$ the beta function of the inflaton self-coupling $\la_\phi$, which we regard as a free parameter in this model independent approximation.
The potential $V(\phi)$ has been projected onto the direction of inflation, i.e. the direction obtained by setting any other scalar field at the minimum of the potential.

\begin{figure}[t!]
\bc
  \includegraphics[width=0.49\textwidth]{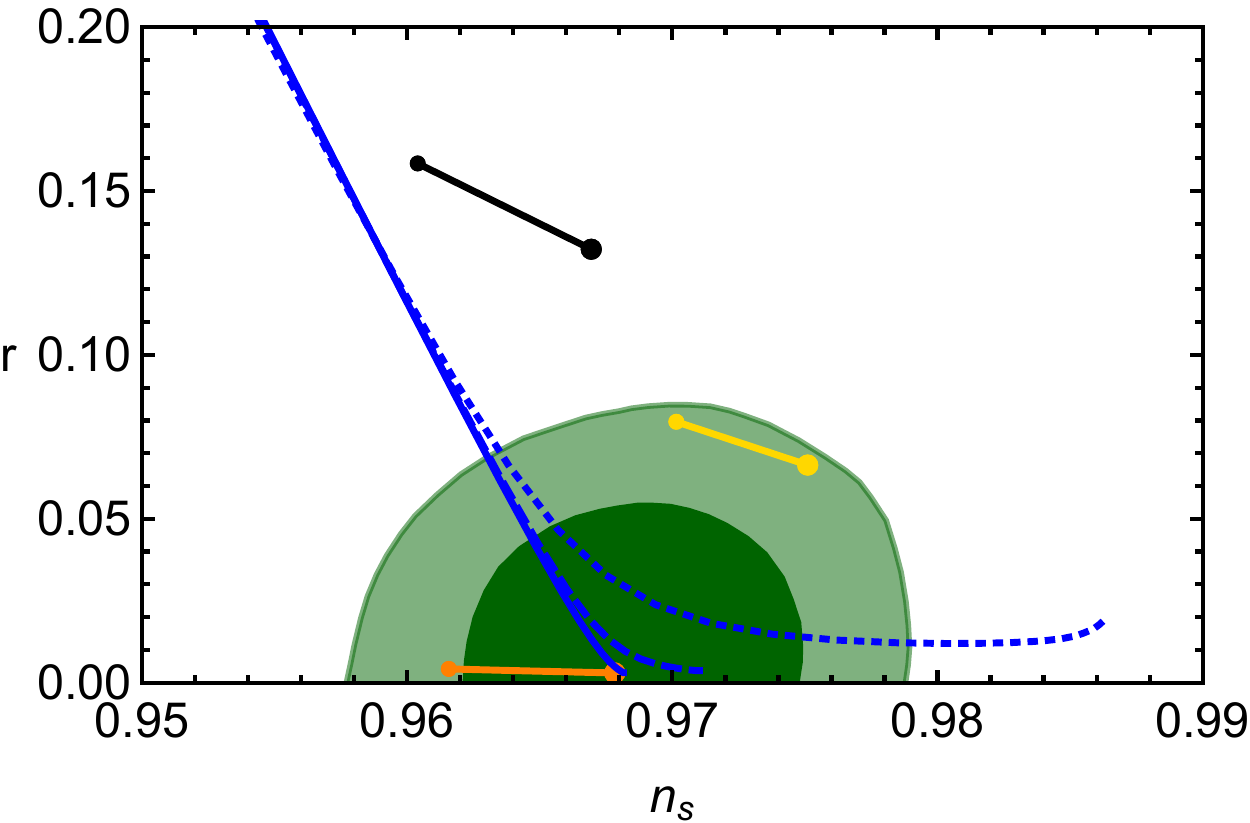}
  \caption{Inflationary results.  Dark and light green areas correspond to the $1\sa$ and $2\sa$ regions from the new BICEP2/Keck Array data \cite{BICEP2new}.  Black/yellow/orange lines represent quadratic/linear/Starobinsky inflation solutions for $50<N_e<60$. The blue lines are the result of our CSI inflation toy model for $N_e=60$ and $\delta(\MP) = 0.001$ (continuos), $0.01$ (dashed), $0.1$ (dotted).
For numerical purposes we took an upper bound $\nmc < e^{8}$.}
  \label{fig:inf:general}
\ec
\end{figure}

Since the slow-roll parameters are independent of the scalar potential normalization, the values of the tensor-scalar ratio, $r$, and the scalar spectral index, $n_s$, depend only on the \emph{relative} quantum correction encapsulated in $\delta$. The results obtained for different choices of this parameter are shown in Fig.~\ref{fig:inf:general}, where we plot in blue the predicted values of $r$ as a function of $n_s$ taking  $N_e=60$ $e$-folds. The dark and light green regions correspond to the $1\sa$ and $2\sa$ confidence intervals from the new BICEP2/Keck Array data \cite{BICEP2new}, whereas the  black, yellow, and orange lines respectively identify the quadratic, linear and Starobinsky solutions for $50<N_e<60$. 
In our analysis we took an upper bound $\nmc < e^{8}$ and safely neglected the quantum corrections to $\nmc$  \cite{Kannike:2015apa,Kannike:2015kda}. In fact, in the same fashion as in eq.~\eqref{eq:lrun}, it is possible to approximate the running of the non-minimal coupling $\nmc$ as
\begin{equation}
  \nmc = \nmc(\MP) + \beta_\nmc(\MP) \ln\left(\frac{\phi}{\MP}\right)
  \label{eq:xirun}
\end{equation}
where $\beta_\nmc$ is the beta function of the non-minimal coupling
\begin{equation}
  16 \pi^2 \beta_\nmc(\phi) \simeq 6 \lambda_\phi(\phi) \left( \xi(\phi) + \frac16 \right)
  \label{eq:betaxi}
\end{equation}
Then, because of the constraint on the amplitude of scalar perturbations \cite{Ade:2015lrj,Ade:2015xua}
\be
A_s = (\PRexp) \times 10^{-9}, \label{eq:As}
\ee
and the $16 \pi^2$ suppression factor, it is $\beta_{\nmc} \ll \nmc$. The quantum corrections are therefore negligible and $\nmc \simeq \nmc(\MP)$. For increasing values of $\xi$, the results move away from the quartic solution towards smaller values of $r$. It is clear that the quantum corrections differentiate the CSI solutions from the Starobinsky ones, favouring higher values of the spectral index $n_s$ in line with the latest measurements \cite{Ade:2015lrj,Ade:2015xua}. For values of $\delta$ and $\nmc$ large enough, the divergence from Starobinsky solution is so strong that CSI solutions even fall outside of the $2\sa$ region.

\begin{table}[t]
\footnotesize
	\centering
\begin{tabular}{|c|c|c|c|c|c|c|c|c|c|}
  \hline			
  $\sin\alpha$ & $m_{\tilde \phi} \text{ (GeV)}$ & $\la_\phi(\MP)$ & $\delta(\MP)$ & $\nmc$ & $r \times 10^3$ & $n_s$ & $\TRH \text{ (GeV)}$ & $\ell\text{ (m)}$\\
  \hline
0.00136 & 0.00974  & 5.89 $\times 10^{-13}$ & 0.00628 & 0.006 & 68.1 & 0.96312 & 0.325  &  8.54 $\times 10^{6}$\\
 0.01 & 0.072  & 1.73 $\times 10^{-9}$ & 0.00628 & 1.83 & 3.68 & 0.96975 & 18.5  &  3.60 $\times 10^{2}$\\
 0.05 & 0.36 & 1.08 $\times 10^{-6}$ & 0.00624 & 47.7 & 3.42  & 0.96994 & 3.12 $\times 10^3$  &  2.53 $\times 10^{-3}$\\
 0.1  & 0.71 & 1.73 $\times 10^{-5}$ & 0.00614 & 191.5 & 3.41  & 0.96991 & 1.13 $\times 10^4$ &  9.71 $\times 10^{-5}$\\
 0.5  & 2.68 & 1.37 $\times 10^{-2}$ & 0.00354 & 4989  & 3.32  & 0.96961 & 2.60  $\times 10^5$   & 4.86 $\times 10^{-8}$\\
 \hline
\end{tabular}
\caption{A set of benchmark points.}
\label{table:results}
\end{table}

\section{A minimal model}
\label{sec:model}
The minimal setup for a CSI model yielding radiative EWSB was first presented in \cite{Allison:2014hna, Allison:2014zya}. It involves a SM singlet, $\phi$, which couples to the SM Higgs boson $h$ via an adimensional portal. The 1-loop scalar potential in eq.~\eqref{eq:Veff} is then approximated as
\begin{equation}
V \simeq  \frac14 \la_h(m_t) h^4 + \frac14 \la_{h\phi}(m_t) h^2 \phi^2 + \frac14 \left[\la_\phi(m_t)+ \frac{\la_{h\phi}(m_t)^2}{8 \pi^2} \ln\left(\frac{\phi}{m_t} \right) \right] \phi^4 \label{eq:VeffH}
\end{equation}
where $h$ is physical Higgs boson and $m_t$ the top-quark mass. In order to avoid cumbersome notation we will henceforth leave the argument ``$(m_t)$'' understood and, as the details of the mechanism have been extensively studied in \cite{Gabrielli:2013hma,Kannike:2014mia,Allison:2014hna, Allison:2014zya}, we will just highlight the main steps. The quantum corrections to $\la_h$ and $\la_{h\phi}$ can be safely neglected in the minimization of the potential as well as for EWSB purposes, hence the only relevant correction is that of $\la_\phi$, which affects the inflationary dynamics. EWSB is obtained when the following equations admit real solutions for the VEVs
\begin{eqnarray}
\lambda _h v_h^2 +  \frac{1}{2} \lambda _{h \phi } v_{\phi }^2  &=& 0 \label{eq:min1}\\
\frac{1}{2} \lambda _{h \phi } v_h^2  +
 \left[ \lambda_\phi + \frac{\lambda _{h \phi }^2}{8 \pi ^2} \left( \frac{1}{4}+  \ln
   \frac{v_\phi }{m_t} \right) \right] v_{\phi }^2  &=& 0 \qquad \label{eq:min2}
\end{eqnarray}
implying that successful EWSB requires $\lambda _{h \phi }<0$.
The mass eigenvalues of the Higgs-like ($\tilde h$) and inflaton-like ($\tilde \phi$) fields, respectively denoted by $ m_{\tilde h}$ and $m_{\tilde \phi}$, as well as the mixing angle $\alpha$ can be computed exactly, and for $|\lambda_{h\phi}|\propto (v_h/v_\phi)^2 \ll 1$ are well approximated by:
\begin{eqnarray}
%
m_{\tilde h}^2 &\simeq & 2 \lambda_h v_h^2 \left(1 + \frac{v_{h}^2}{v_\phi^2}\right) \label{eq:mh}\\
%
m_{\tilde \phi}^2 &\simeq & \frac{\lambda^2_h v_h^4}{2 \pi ^2 v_\phi^2} \label{eq:mphi}\\
%
\sin\alpha & \simeq & \frac{v_h}{v_\phi}.
\end{eqnarray}
The current collider experiment cast an upper bound on the mixing angle: $\sin \alpha \lesssim 0.51$ at 95\% CL \cite{Cheung:2015dta}. From eq. (\ref{eq:VeffH}) it is clear that the model has originally three free parameters, the couplings in the scalar potential, which can be replaced via eqs (\ref{eq:min1}), (\ref{eq:min2}) and (\ref{eq:mh}) with the following set of phenomenological paramaters: $v_h$, $m_h$, $\sin \alpha$. As the EW precision measurements and the LHC result set $v_h \simeq 246$ GeV and $m_h \simeq 125.09$ GeV \cite{Aad:2015zhl}, $\sin \alpha$ is the only parameter that can still be adjusted. By varying this quantity within the allowed range we then derive a set of corresponding inflaton masses, presented in the first two columns of Table \ref{table:results}. Clearly the upper bound on $\sin\alpha$ induces an upper bound on the inflaton mass $m_{\tilde \phi} \lesssim 2.68$ GeV.

%

\subsection{Inflation and reheating}
\label{subsec:inflation}
In order to compute the necessary inflationary observables we must first derive the values of the involved couplings at the inflationary scale by solving the RGEs of the model. The full set of RGEs, found in \cite{Kannike:2015apa,Kannike:2015kda}, was solved numerically by running from the top-mass scale $m_t$ to one orders of magnitude above the Planck scale\footnote{A careful reader might notice that our setup suffers of the Higgs vacuum stability problem \cite{Buttazzo:2013uya}. In the following we assume that  BSM physics  (for example dark matter \cite{Gabrielli:2013hma,Kadastik:2011aa})  solves the issue without directly coupling to the inflaton, so that our computations still hold. A full analysis of the possible interplay between BSM and inflation is postponed to a future work \cite{toappear}.}, ensuring the perturbativity of the model at all these scales. For the purpose of our discussion, however, we consider here the approximation used in Section \ref{sec:cwi} with $\beta_{\la_\phi} \simeq \frac{\la^2_{h\phi}}{8\pi^2}$. We also emphasize that the overall normalization of the scalar potential is now point by point set by the chosen values of the Higgs boson VEV, mass, and the sine of the mixing angle. Hence, we seek values of the non-minimal coupling $\nmc$ such that the constraint on the amplitude of scalar perturbations in eq. (\ref{eq:As}) is satisfied within two standard deviations. The corresponding results are given in Table \ref{table:results} for $N_e=60$. In our model we typically obtain $0.001< \delta (\MP)<0.01$, therefore the solutions lie between the continuos
and dashed
blue lines of Fig.~\ref{fig:inf:general} and are clearly separated from the Starobinsky one. Moreover, notice that the the non-minimal coupling to gravity decreases as we consider smaller values for the mixing angle. Inflationary constraints will therefore impose a lower bound on the latter:
at 95\% confidence level $r \lesssim 6.81 \times 10^{-2}$ \cite{BICEP2new}, implying that $\sin\alpha \gtrsim 1.36 \times 10^{-3}$ and, therefore, $ m_{\tilde{\phi}} \gtrsim 9.74$ MeV  and $n_s \gtrsim 0.963 $.

Given the very low inflaton mass resulting, reheating can only proceed via the decay into electrons, muons, photons and pions through the mixing with the Higgs boson. The reheating temperature is estimated in
\begin{equation}
  \TRH = \left( \frac{90}{g_{*} \pi^{2}} \right)^{\frac{1}{4}} \sqrt{\Gamma_{\tilde \phi} \MP},
\end{equation}
where $g_{*} \simeq 106.75$ is the number of relativistic degrees of freedom and
\begin{equation}
  \Gamma_{\tilde \phi} \simeq \sin^2 \alpha \ \Gamma_{h}(m_{\tilde \phi}),
\end{equation}
is the total decay width of the inflaton.
Here  $\Gamma_{h}(m_{\tilde \phi})$ is the SM Higgs decay width, computed at the mass $m_{\tilde \phi}$. The results obtained for the reheating temperature, reported in Table~\ref{table:results} for the considered benchmark points, allow for baryon asymmetry production via resonant leptogenesis if $\sin\alpha \gtrsim 0.1$. The lower bound $\sin\alpha \gtrsim 1.36 \times 10^{-3}$ implies a lower bound on the reheating temperature $\TRH \gtrsim 325$ MeV, which respects the experimental bound $\TRH \gtrsim 4.7$ MeV \cite{Dai:2014jja,Munoz:2014eqa,deSalas:2015glj}.

\subsection{Collider phenomenology}
\label{sec:LHC}
Another interesting signature of this model is the potential observation of inflaton decays at the LHC or ILC. Because of the light mass and the small mixing, the inflaton may be long-living and travel enough to generate a displaced vertex in the detector, or even decay on the outside of the latter. The discriminant is the relativistic boost of the inflaton with respect to the lab frame, resulting from its production mainly via $\tilde h \to \tilde \phi\tilde \phi$. For simplicity we opted to estimate this quantity at the Higgs resonance, hence each generated inflaton carries an energy $E_{\tilde \phi} = m_{\tilde h}/2$.
The results for the corresponding decay length $\ell$ are presented in the last column of Table \ref{table:results}; as expected, smaller mixing angles yield increasingly displaced vertices. For $\TRH\sim\mathcal{O}$(TeV), our model predicts decay length that are in principle appreciable with the current technology. For higher values of this parameter our framework would instead contribute to the apparent decay of the Higgs boson into four electrons or muons.

\section{Conclusions}
\label{sec:end}
We discussed the most minimal CSI extension of the SM able to account for successful EWSB and inflation. Such a model involves an extra scalar singlet that plays the r\^ole of the inflaton and relies on a non-minimal coupling of the latter to gravity. EW bounds reduce the available free parameters to the Higgs-inflaton mixing angle, $\alpha$. Inflationary data \cite{BICEP2new} set the upper bound for the tensor-to-scalar ratio to $r \simeq 0.068$, which induces the lower bounds $\sin\alpha \gtrsim 1.4 \times 10^{-3}$ and  $\TRH \gtrsim 325$ MeV for the reheating temperature. The present upper bound on $\sin\alpha$ due to collider experiments sets instead the maximal inflaton mass to $m_{\tilde \phi} \lesssim 2.68$ GeV. We obtain a lower limit for $r$ of about $3 \times 10^{-3}$, which along with a scalar spectral index $n_s \simeq 0.97$ makes our solutions diverge from the Starobinsky one. For small values of the mixing angle, $\sin\alpha\lesssim 0.01$ as favoured by LHC data \cite{Cheung:2015dta}, the inflaton quartic coupling is naturally small and a non-minimal coupling to gravity $\xi \lesssim 1$ is enough to satisfy the constraint on $A_s$ \cite{Ade:2015lrj,Ade:2015xua}. In our setup the inflaton may be enough long-living to travel macroscopic distances, generating a displaced vertex in collider experiments or even decaying outside of the detectors. Future measurements with better  sensitivity to $r$ and $n_s$ may test our scenario.

\section*{Acknowledgments}
This work was supported by the Estonian Research Council grants
PUTJD110 , PUT1026,  IUT23-6 and through the ERDF CoE program.


\bibliographystyle{JHEP}
\bibliography{citations}

\end{document}